\documentclass[10pt,sigconf,letterpaper,nonacm]{acmart}
\usepackage{listings}
\usepackage{array}

\begin{document}

\title{Acila: Attaching Identities of Workloads for Efficient Packet Classification in a Cloud Data Center Network}

\author{Kentaro OHNISHI}
\email{ohnishi@net.ist.i.kyoto-u.ac.jp}
\affiliation{%
  \institution{Kyoto University}
  \streetaddress{Yoshida-Honmachi}
  \city{Sakyo}
  \state{Kyoto}
  \country{Japan}
  \postcode{606-8501}
}

\author{Daisuke Kotani}
\email{kotani@media.kyoto-u.ac.jp}
\orcid{0000-0003-4305-8379}
\affiliation{%
  \institution{Kyoto University}
  \streetaddress{Yoshida-Honmachi}
  \city{Sakyo}
  \state{Kyoto}
  \country{Japan}
  \postcode{606-8501}
}

\author{Hirofumi Ichihara}
\email{hirofumi.ichihara@linecorp.com}
\affiliation{%
	\institution{LINE Corporation}
	\streetaddress{4-1-6 Shinjuku}
	\city{Shinjuku}
	\state{Tokyo}
	\country{Japan}
	\postcode{160-0022}
}

\author{Yohei Kanemaru}
\email{yohei.kanemaru@linecorp.com}
\affiliation{%
	\institution{LINE Corporation}
	\streetaddress{4-1-6 Shinjuku}
	\city{Shinjuku}
	\state{Tokyo}
	\country{Japan}
	\postcode{160-0022}
}

\author{Yasuo Okabe}
\email{okabe@i.kyoto-u.ac.jp}
\orcid{0000-0003-0825-2256}
\affiliation{%
  \institution{Kyoto University}
  \streetaddress{Yoshida-Honmachi}
  \city{Sakyo}
  \state{Kyoto}
  \country{Japan}
  \postcode{606-8501}
}


\begin{abstract}
IP addresses and port numbers (network based identifiers hereafter) in packets are two major identifiers for network devices to identify systems and roles of hosts sending and receiving packets for access control lists, priority control, etc.
However, in modern system design on cloud, such as microservices architecture, network based identifiers are inefficient for network devices to identify systems and roles of hosts.
This is because, due to autoscaling and automatic deployment of new software, many VMs and containers consisting of the system (workload hereafter) are frequently created and deleted on servers whose resources are available, and network based identifiers are assigned based on servers where containers and VMs are running.
In this paper, we propose a new system, Acila, to classify packets based on the identity of a workload at network devices, by marking packets with the necessary information extracted from the identity that usually stored in orchestrators or controllers.
We then implement Acila and show that packet filtering and priority control can be implemented with Acila, and entries for them with Acila is more efficient than conventional network based identifiers approach, with little overhead on performance.
\end{abstract}


%\begin{CCSXML}
%<ccs2012>
%<concept>
%<concept_id>10003033.10003099.10003104</concept_id>
%<concept_desc>Networks~Network management</concept_desc>
%<concept_significance>500</concept_significance>
%</concept>
%<concept>
%<concept_id>10003033.10003099.10003037</concept_id>
%<concept_desc>Networks~Naming and addressing</concept_desc>
%<concept_significance>300</concept_significance>
%</concept>
%<concept>
%<concept_id>10003033.10003099.10003100</concept_id>
%<concept_desc>Networks~Cloud computing</concept_desc>
%<concept_significance>300</concept_significance>
%</concept>
%</ccs2012>
%\end{CCSXML}
%
%\ccsdesc[500]{Networks~Network management}
%\ccsdesc[300]{Networks~Naming and addressing}
%\ccsdesc[300]{Networks~Cloud computing}

\keywords{Cloud Network, Identity, Workload}

\maketitle

\section{Introduction}
\label{sec:hajimeni}

In a cloud platform, various types of software, such as VMs, containers and UNIX processes, run and work closely together to form a scalable software development and execution platform. \emph{Workload}~\cite{SPIFFE} is a concept for managing such various types of software comprehensively.
VMs and containers are examples of the workload.

Various packet processing should be applied at network devices like routers and switches to secure the workload, such as packet filtering for discarding packets from unintended workloads, and bandwidth and priority control for surely transmitting important packets.
Conventionally, network based identifiers such as IP addresses and port numbers are used to classify packets in network devices for identifying who is communicating, because a prefix of IP addresses represents a group of similar workloads, and port numbers are known.
However, for inter-workload communications in a cloud platform, it is difficult to widely use network based identifiers for this purpose due to the following two reasons.

First, modern system design on cloud, such as microservice architecture\cite{Microservices}, requires a large number and frequent updates of entries for classifying packets in network devices.
Workloads of a system are frequently created and deleted due to autoscaling, deploying new software, etc, and each workload is created on servers where computing resources are available.
Use of containers accelerates the speed of creation and deletion of workloads.
In addition, IP addresses on each server are assigned based on the location of the server\cite{FBBGP}\cite{LINEDC} on the network, and some of them will be reassigned to the workloads on the server, so the prefix of IP addresses does not represent a group of similar workloads.
Furthermore, more and more types of workloads will be deployed due to the nature of microservice architecture, which divides a large application into multiple loosely coupled applications that communicate with each other.

Second, granularity of network based identifiers is sometimes too low to identify workloads at network devices.
Small workloads such as UNIX processes or containers may share one IP address assigned to VM and use random source port numbers, thus workloads working as clients cannot be identified by IP addresses and port numbers.

At the same time, we have to consider the fact that many legacy systems, which are not orchestrated by the cloud controller, are sometimes essential for systems on the cloud when the legacy systems cannot be easily replaced.
This means network devices have to classify packets to/from many legacy systems and handle packets properly.

Overlay networks enable to assign IP addresses in the same prefix to multiple similar workloads.
However, overlay networks incur additional overhead on the design of the system, e.g., network devices to process packets by access control lists (ACLs) must be placed on the overlay networks.
This may hinder flexible system configuration changes, which is one of the benefits of the cloud.
Service Mesh~\cite{ServiceMesh} represented by Istio~\cite{Istio} is a middleware that performs various processing such as access control to application level messages.
This is achieved by intercepting and modifying packets at a proxy running along with each application, but Istio cannot prevent unintended packets processed at the proxy itself or OS.

This paper proposes Acila, a system to directly mark packets with the necessary information extracted from the identity of workloads, which mainly comes from orchestrators or controllers in the cloud, so that network devices can classify packets based on the identity of a workload.
We assign a \emph{Service} to workloads whose packets are processed in the same way in the network, which is derived from the identity of the workloads.
Then, we add an identifier of the service (called SACL ID) to packets, and network devices classify packets based on SACL IDs for ACLs, bandwidth and priority control, etc.
In this way, the number of entries can be decreased and there is no need to make any changes to entries even if a set of workloads associated with the same Service is changed.

We design and implement Acila, and we show that Acila can be applied to various use cases with two examples, packet filtering and priority control.
The evaluation shows that mechanisms with Acila greatly reduce the burden of managing entries for classifying packets with little overhead on packet processing performance.

The contribution of this paper is summarized as follows.

\begin{itemize}
\item We design Acila, a system for Service-based packet classification by marking packets with information extracted from the identity of workloads that send and receive packets.
By adding both Services of sender and receivers to packets, Acila can support not only containers and virtual machines, but also legacy systems through gateways.

\item We demonstrate Acila has various applications with two examples, access control lists whose processing should be applied at somewhere on the path to the destination, and priority control whose processing should be applied at everywhere to the destination.

\item We show that Acila greatly reduces the burden of managing entries for classifying packets by calculating necessary entries to be installed on the network devices, and comparing with the conventional network based identifier approach. The performance overhead caused by introducing Acila is small.

\end{itemize}

\section{Related Works}

\textbf{Protocols} There are many proposals to separate host identifiers and location identifiers to identify hosts regardless of locations where hosts are, such as HIP~\cite{HIP}, Mobile IP~\cite{MobileIP} and LISP~\cite{LISP}.
In our scenario, identifiers in each workload level are too detail for packet processing like packet filtering, bandwidth control and priority control, and the number of entries and update frequencies at network devices are still high.

We can build an overlay network using protocols like IPIP~\cite{IPIP} and VXLAN~\cite{VXLAN}, so that IP addresses with the same prefix are assigned to similar workloads, and entries can be aggregated by the prefixes in a cloud platform.
On the other hand, entry management by prefixes may hinder flexible system configuration changes, which is one of the benefits of cloud platforms.

Guichard et al.~\cite{SRHCLS} published a concept which inserts source and destination `Classes' which is similar to our `Services' into SRH~\cite{SRv6}, a packet format for segment routing in IPv6, so that network devices such as firewalls can identify Classes and reduce the number of entries.
Their purpose is similar to us, but the published concept is still in the early stage, and there remain many undiscussed topics, such as detailed packet format, methods to assign Classes, and evaluation. Also, this protocol is built on SRH so segment routing needs to be deployed to the network.

Tools for data plane programmability like P4~\cite{P4} allows operators to specify how programmable network devices like Tofino~\cite{Tofino} process packets.
Thanks to this, it is now more realistic to process packets having our own packet format at line rate.
We can utilize this tool for real deployment of the proposed method.

\textbf{Access control for microservices on cloud}
SPIFFE~\cite{SPIFFE} is a framework that manages the identities of workloads in a cloud platform.
This authenticates various types of workloads, such as UNIX processes and containers, and gives a certificate of the corresponding identity to the workloads.
SPIFFE assumes that workloads perform mutual authentication with endpoints by mTLS using its certificates, and Istio~\cite{Istio} implements access control in the same idea.
In this approach, other than payload within mTLS, such as attacks to the kernel and codes to handle mTLS itself, is not protected.

Cilium~\cite{Cilium} executes packet filtering and tracing based on the identity of workloads.
Cilium adds identifier to packets about the identity of source workload at the source node, and applies actions to packets at the destination node.
eZTrust~\cite{eZTrust} attaches similar identifiers to packets.
The identifiers can be mapped to more information than Cilium's one.
Such information includes process ID, UID and others captured by in-kernel tracers, and is stored in the centralized context map.
Cilium and eZTrust attaches some identifiers related to the source workload only.
However, for network devices to apply ACLs, priority and bandwidth control, etc, information related to the destination is also required to identify the type and importance of the traffic.
We propose to add information about the identity of both workloads, at the source and the destination, to packets so that we can perform such processing within the network as well as end hosts.

\section{A Premise of Cloud Platform}
\subsection{Workloads and Labels}

\label{sec:workload}

A piece of running software that provides a single function is called a workload~\cite{SPIFFE}.
For example, in IaaS (Infrastructure as a Service), a VM provided to users from the cloud administrator or a UNIX process or a container in the VM can be a workload.
In CaaS (Container as a Service), a pod in Kubernetes or a container can be a workload.

As shown in Figure~\ref{fig:register_workloads}, a workload can be identified from outside of the workload by the IP address, the listening port number, UNIX process information (e.g.~UID), etc., and a combination of these.

Administrators of workloads (or cloud users) assign labels to the workloads for management purposes.
Labels contain metadata about the workload, e.g., customer, service name, role and version, and each label consists of a key and a value.
Typical cloud platforms already have such kind of labels in their cloud orchestrators as metadata of the workload, assigned by cloud administrators or the users.
For example, Kubernetes~\cite{Kubernetes} has metadata of workloads in the form of a set of key and its values.
When such kind of information is not available or too coarse, we assume that further information is provided directly from the cloud administrators and the users.

\subsection{Trust Model in a Cloud Platform}

\begin{figure}[ht]
	\begin{center}
		\includegraphics[width=\linewidth]{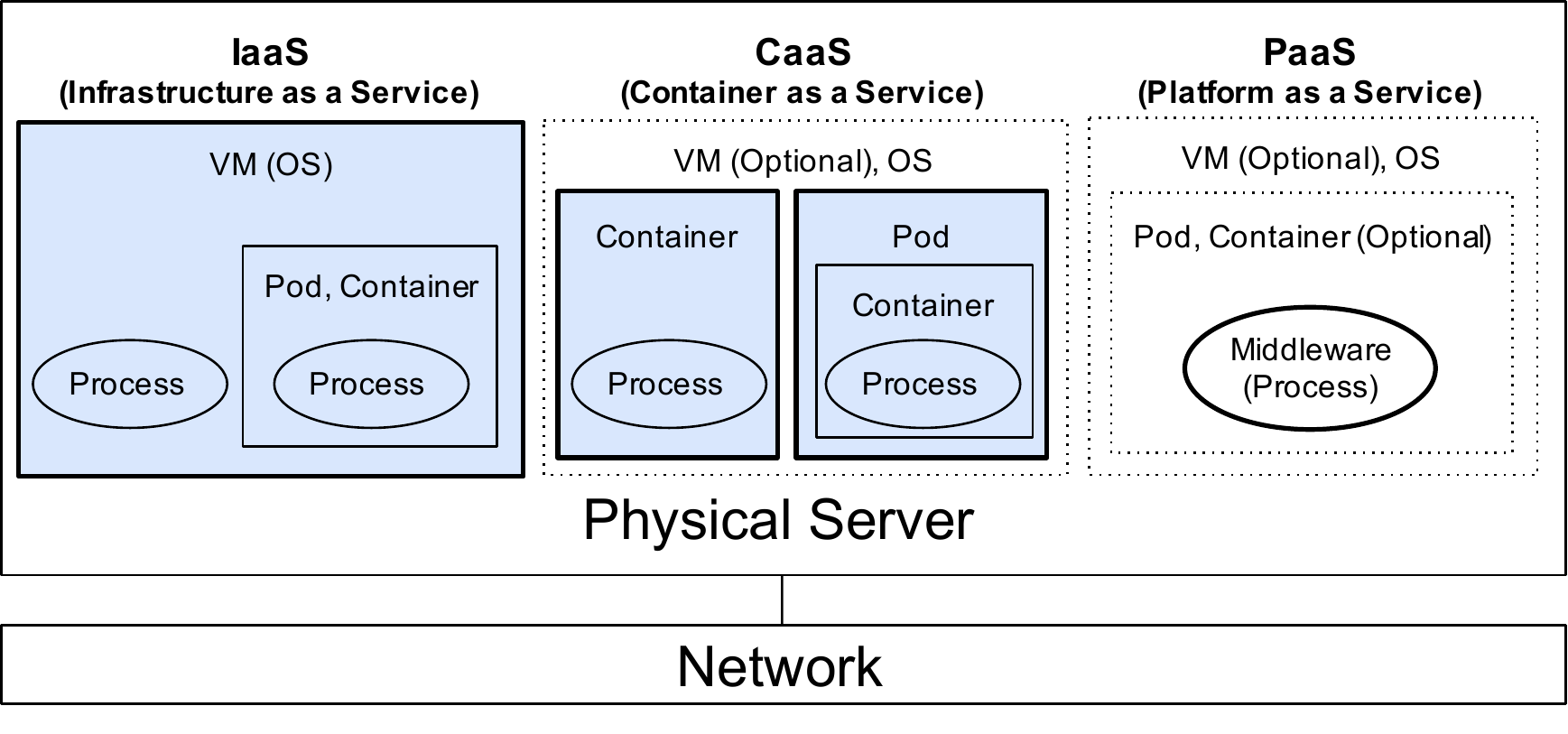}
		\caption{Administrators in a cloud. Blue indicates administrative privileges is in the user while others in the administrator of the cloud.}
		\label{fig:cloud}
	\end{center}
\end{figure}

As shown in Section~\ref{sec:workload} and Figure~\ref{fig:cloud}, workloads are categorized to several types according to administration privileges that the cloud administrator allows a user to have.
In this paper, entities managed by the cloud administrator are treated as trustworthy, while those managed by cloud users are not treated as completely trustworthy.

For example, in IaaS, the cloud administrator deploys a VM and provides it to cloud users, and the users manage the VM.
The cloud administrator can confirm that packets sent to and received from the VM are truly done by the VM, but payloads of the packets cannot be trusted because the cloud user can change the payloads of the packets from the VM\@.
Therefore, there is no reliable way for the cloud administrator to identify workloads inside VMs such as UNIX processes and containers, and they have to rely on the users to identify them.
In CaaS, the cloud administrator can reliably identify workloads running as containers because the infrastructure that runs containers are managed by the cloud administrators.

\subsection{Target Packet Processing}
\label{sec:target_types}

The proposed method targets packet processing that requires more metadata about workloads than network based identifiers to configure it.
Such metadata includes owners of the workloads and roles in the system.
The administrators use metadata to express policies indicating how to handle packets to/from workloads.
In conventional network devices, administrators translate such policies into rules using network based identifiers, and install such rules and intended actions to be applied to packets into the devices.

For example when an administrator tries to configure that database servers are accessible only from application servers, the administrator translates the database servers and the application servers into a set of IP addresses, and installs packet filtering rules using IP addresses into network devices.
Other examples include prioritizing traffic related to workloads whose response time is sensitive to the overall system performance, limiting bandwidth consumed by workloads that handle background jobs, etc.

The proposed method tries to remove the use of network based identifiers from such rules and to provide alternative identifiers, given the nature of the workloads on the cloud where usage of resources is elastically changed.

\section{Acila: Assign Workload Identity Based Identifiers to Packets}

\begin{figure}[ht]
	\begin{center}
		\includegraphics[width=0.7\linewidth]{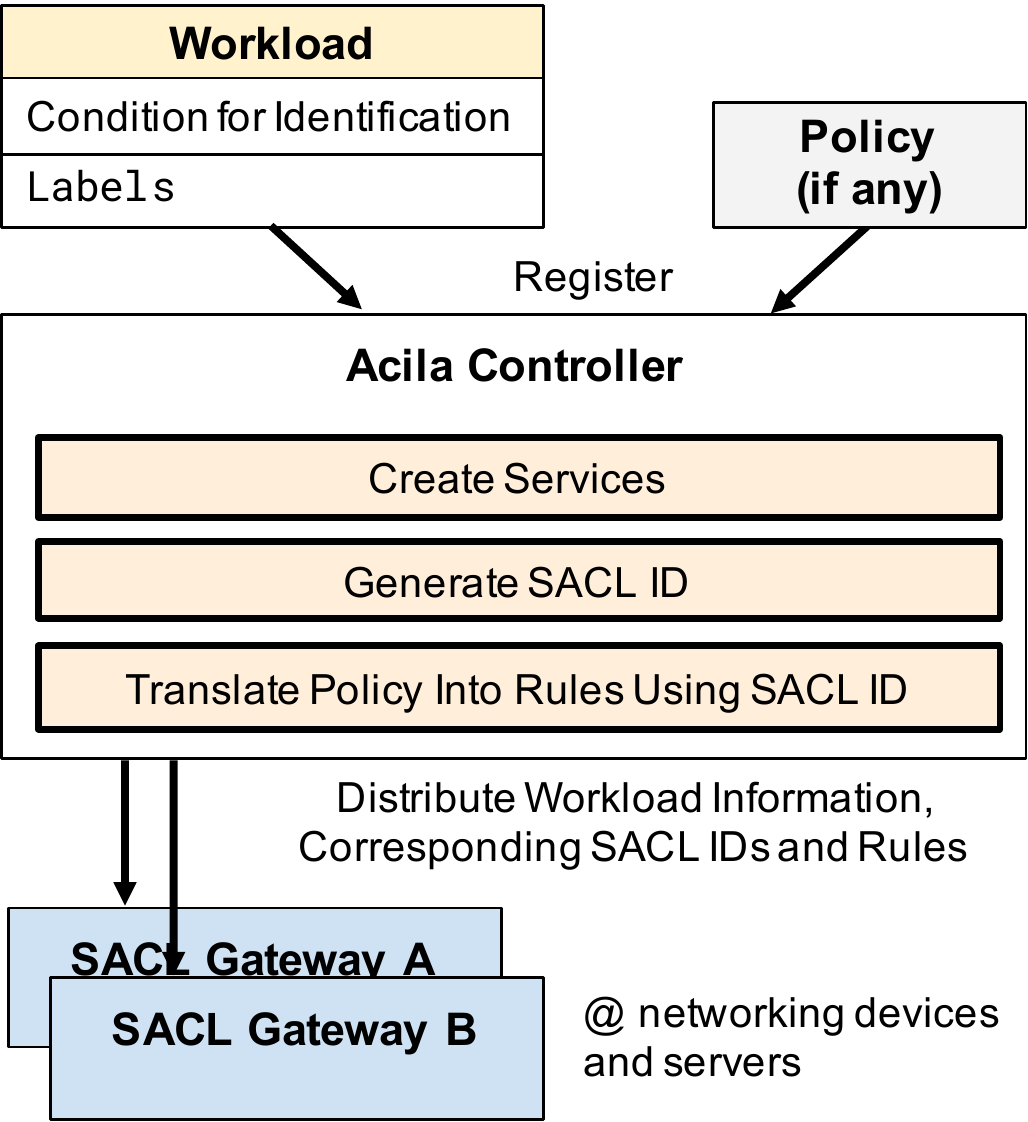}
		\caption{Acila overview}
		\label{fig:acila_overview}
	\end{center}
\end{figure}

Figure~\ref{fig:acila_overview} shows an overview of the proposed system, Acila.
Acila Controller receives workload information and policies (if any) as input from cloud orchestrators and administrators.
Acila Controller creates Services and SACL ID corresponding to each Service, and translates policies into rules using SACL ID.
Then, the controller distributes workload information, SACL IDs and rules to SACL Gateways in network devices and servers, which act as a data plane handling packets with SACL ID.

This section explains handling SACL IDs including attaching them to packets.
Policy is covered at Section~\ref{sec:policy}.

\label{sec:assign_sevice_to_w}

\subsection{Service}

The identity of a workload is the information that characterizes the workload, which includes not only network based identifiers but also application types, application characteristics, versions, servers deployed on, etc., and such information is expressed as labels.
This information varies according to workloads, but not all information is used to express policies by the administrators.
Rather, the policies use a subset of labels to select workloads whose packets are processed in the same way.
\emph{Service} represents a subset of labels required to classify packets that the same actions are applied.
Then we associate each workload with a Service based on its labels.

\begin{figure}[ht]
	\begin{center}
		\includegraphics[width=\linewidth]{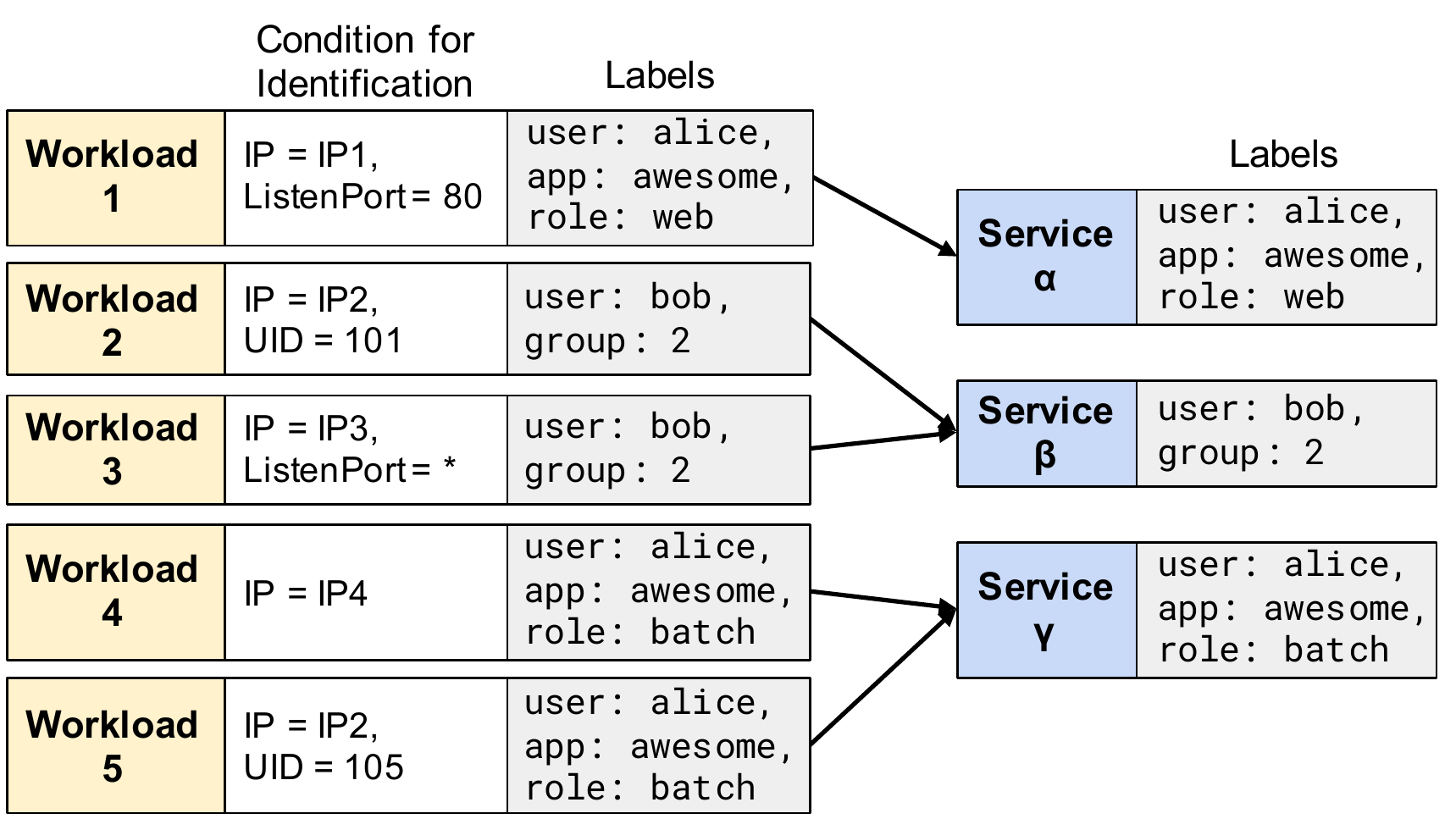}
		\caption{Workloads, Labels and Services}
		\label{fig:register_workloads}
	\end{center}
\end{figure}

For example in Figure~\ref{fig:register_workloads}, given the same actions should be applied to the workloads whose labels are \texttt{\{user:~bob, group:~2\}}, we create Service $\beta$ that has the same labels, and then assign workload 2 and workload 3 to Service $\beta$.

Network devices have to identify Services corresponding to the source and destination workloads.
To do this, we add an identifier of the Service called \emph{SACL ID} to packets. SACL ID is unique among all Services.

For simplicity, we assume that the labels associated with a workload and those associated with a Service have a one-to-one correspondence.
Workloads with the same labels belong to the same Service.
An example of the correspondence is shown in Figure~\ref{fig:register_workloads}.

The creation and assignments of Services are done in Acila Controller, which is an application that manages various data used in Acila.
Once a workload is created or updated, such event is sent to Acila Controller with its labels.
At this time, if a Service with the same labels already exists, the Service is assigned to the workload.
Otherwise, Acila Controller creates a new Service corresponding to the labels and assigns it to the workload.

\subsection{Adding SACL ID to Packets}
\label{sec:add_saclid}

\begin{figure}[ht]
	\begin{center}
		\includegraphics[width=\linewidth]{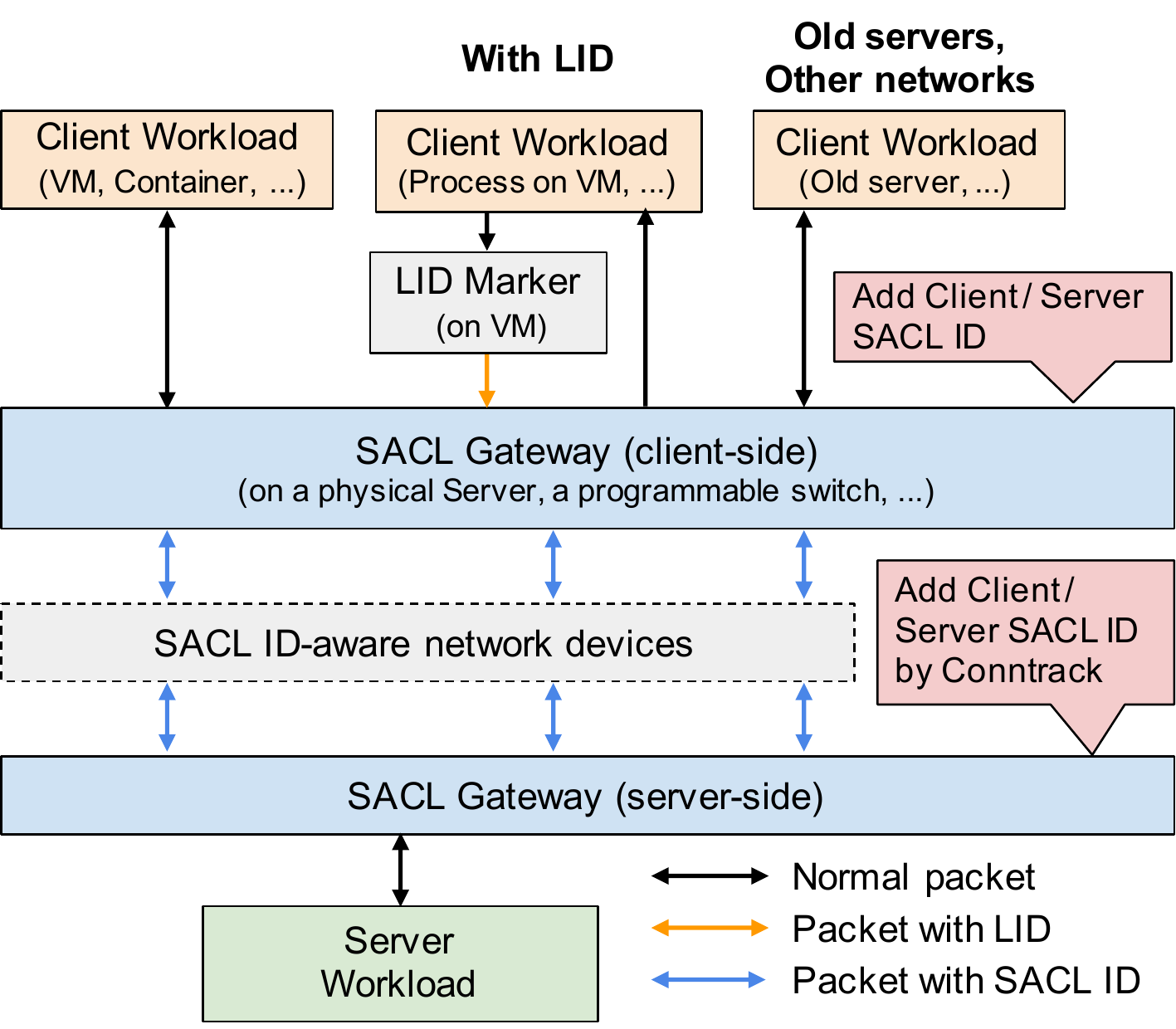}
		\caption{Overview of adding SACL ID to packets}
		\label{fig:how-to-attach-sacl-id}
	\end{center}
\end{figure}

An overview of adding SACL ID to packets sent from workloads is shown in Figure~\ref{fig:how-to-attach-sacl-id}.

A workload is divided into a Client Workload (on the client-side) and a Server Workload (on the server-side) since the method of identifying workloads and adding SACL IDs greatly differs.
Corresponding Services are called a Client Service and a Server Service.
Acila Controller regards all workloads as Client Workloads.
When any listen port number is specified in the condition for identification of workload, the controller recognizes that Server Workloads are also registered, assuming that it also acts as a server.
In this case, the listen port number is excluded in the condition for identification.

\emph{SACL Gateway} is an entity that adds and removes SACL IDs to/from packets.
The SACL Gateway is logically in the middle of Client/Server Workloads and the rest of the network (called Client/Server Workloads are under the SACL Gateway), and has information of the Client Workloads and the Server Workloads that is necessary for adding SACL IDs to packets.
For example, in the case of packet filtering, SACL Gateway holds the information of the Server Workloads that are allowed to communicate from the Client Workloads under the SACL Gateway.

The processing of SACL Gateway is attached to the interface where each workload is connected and the cloud administrator manages.
For example, when a VM is a workload, SACL Gateway will intercept and modify packets on the TAP interface through which the hypervisor host forward packets to/from the VM or the physical interface of the hypervisor host.

The reason is available information to identify workloads and trustworthiness of SACL IDs in packets.
When attaching the SACL Gateway on an interface near the workload, rich information that can only be known near the Client Workload, such as the Service IP in Kubernetes, is available to identify the Server Workload from the destination of a packet.
This leads to precise identification of the Server Workload with a smaller number of entries in the SACL Gateway.
On the other hand, when the SACL Gateway is installed in the area managed by the user, such as inside VMs or containers, the user may forge the SACL ID in packets, and the cloud administrator cannot trust the SACL ID in packets and additional mechanism to verify the SACL ID is necessary.

\label{sec:sacl_id_validate}

For devices that cannot implement SACL Gateway in it or external networks, network devices that support the extension of the SACL ID (e.g.~programmable switches) should be installed and run as SACL Gateway.

\textbf{SACL Gateway Processing} When packets are sent from a Client Workload to a Server Workload, the Client SACL ID is added by identifying the Client Workload based on the source IP address, and the Server SACL ID is added by identifying the Server Workload based on the destination IP address and port number.

For the reverse direction, from a Server Workload to a Client Workload, it is difficult to identify the Client Workload from the packet because the port number used by the client is not known in advance. To solve the problem, the server-side SACL Gateway performs Connection Tracking (Conntrack)~\cite{Conntrack} to find the Client SACL ID from network based identifiers in packets. Conntrack remembers correspondences between the network based identifiers and the SACL ID of active connections between workloads.

SACL Gateway removes SACL ID from arriving packets and forwards them to the workloads.

\textbf{LID} Some Client Workloads, such UNIX processes in a VM managed by a user, cannot be identified only by the information that the cloud administrator has.
The port number is not sufficient since the client uses random source port numbers.
A connection is not stored in Conntrack when the workload starts to send packets.

To identify such workloads, we have to use internal information in the computing resource managed by the user.
For example, when a UNIX process in a VM managed by the user is defined as a workload, only the VM can identify which process sends packets.
However, we do not want to put SACL Gateway into the VM because the SACL ID will become easily forged and untrustworthy.

For the above reason, we design a mechanism to partially delegate the trust of a SACL ID associated with a workload to the cloud user who manages the workload, in a way that only affects the user.
This mechanism conveys the information for identifying the workload from the computing resource managed by the user to the SACL Gateway.
The cloud administrator adds the SACL ID at the SACL Gateway using this information.

The computing resource that can identify the workload between the workload and the SACL Gateway, such as VM in the previous example, adds a temporary identifier of workload to the packets and conveys it to the SACL Gateway.
The identifier is called \emph{LID} (Local ID) and \emph{LID Marker} is installed into the computing resources to add the LID to the packets.
SACL Gateway prevents a malicious LID Marker from impersonating a workload that is not under the LID Marker by mapping a pair of LID Marker and LID to a Client Workload.

This mechanism does not prevent the LID Marker from impersonating another Client Workload under the same LID Marker.
Users can use LID Marker for finer-grained workload identification that cannot be done by the cloud administrator alone.
Instead, the management of the LID Marker and the trustworthiness of LID are left to the user.

\subsubsection{Packet Format}

\begin{figure}[ht]
	\begin{center}
		\includegraphics[width=\linewidth]{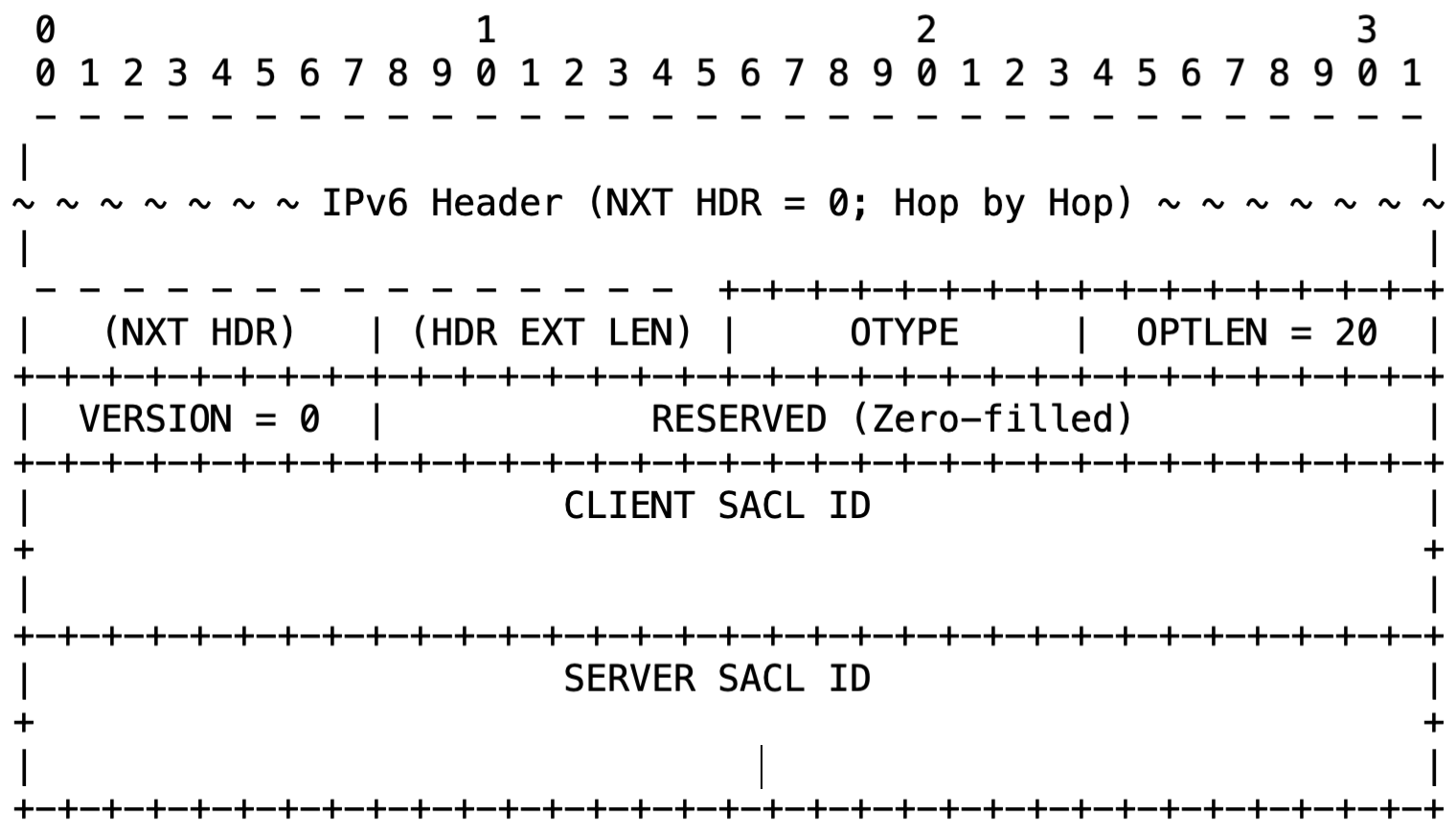}
		\caption{Packet format for transmitting SACL ID}
		\label{fig:packet_format}
	\end{center}
\end{figure}

We use TLV (Type-length-value) in Hop-by-Hop Options of IPv6~\cite{IPV6} as the packet format for transmitting SACL IDs. This is because IPv6 is being adopted in large-scale cloud data center networks, due to limited private IPv4 address space, and the following two reasons. Figure~\ref{fig:packet_format} shows the packet format whose SACL ID is 64 bit long.

\textbf{Flexibility} A field with a fixed length such as IPv6 flow labels~\cite{IPV6} is not suitable since the bit length of SACL ID will be limited by such fields. TLV has a variable length so we can design SACL IDs with sufficient length according to the scale of a cloud platform and make future extensions easy.

\textbf{Compatibility}
It is possible to reuse header fields defined by existing protocols such as VXLAN~\cite{VXLAN}, Geneve~\cite{Geneve} and MPLS~\cite{MPLS} for adding SACL IDs to packets.
However, network devices may perform unexpected interpretation and processing by values in header fields reused for different purposes.

Since IPv6 Hop-by-Hop Option is standardized as a part of IPv6, network devices can treat it as a normal IPv6 packet.
Also, in the Hop-by-Hop Option (TLV), we can specify the action of the devices when the devices do not support the Type, with the first 2 bits of the Type.
If the value is \texttt{00}, the device will `skip over this option and continue processing the header~\cite{IPV6}.
So even if the device is unaware of SACL ID, the device just ignores the SACL ID and continues processing, and the packet will not be discarded as an unknown protocol.

\section{Realizing Target Packet Processing Types with Acila}
\label{sec:policy}

This section explains how we can apply the proposed method to targeted packet processing types in Section~\ref{sec:target_types}, packet filtering and priority control as examples.

An entry in a network device is in the form of a set of \texttt{(Client SACL ID, Server SACL ID, Value (Option))}.
Value indicates additional information such as priority.

\subsection{Policy Management and Distribution}

\emph{Policy} is a set of rules to select a group of Client Workloads and a group of Server Workloads based on labels, and to specify actions or Values for actions applied to packets.
Policy is basically registered and managed by cloud users.
ABAC (Attribute-Based Access Control)~\cite{ABAC} is used for this selection because ABAC has a high affinity with labels.

The users register a policy to select workloads based on their labels without being aware of the Services.
Acila Controller interprets the policy to select appropriate Services, which is generated from the labels of the workloads.

\begin{figure}[!ht]
	\begin{center}
		\includegraphics[width=\linewidth]{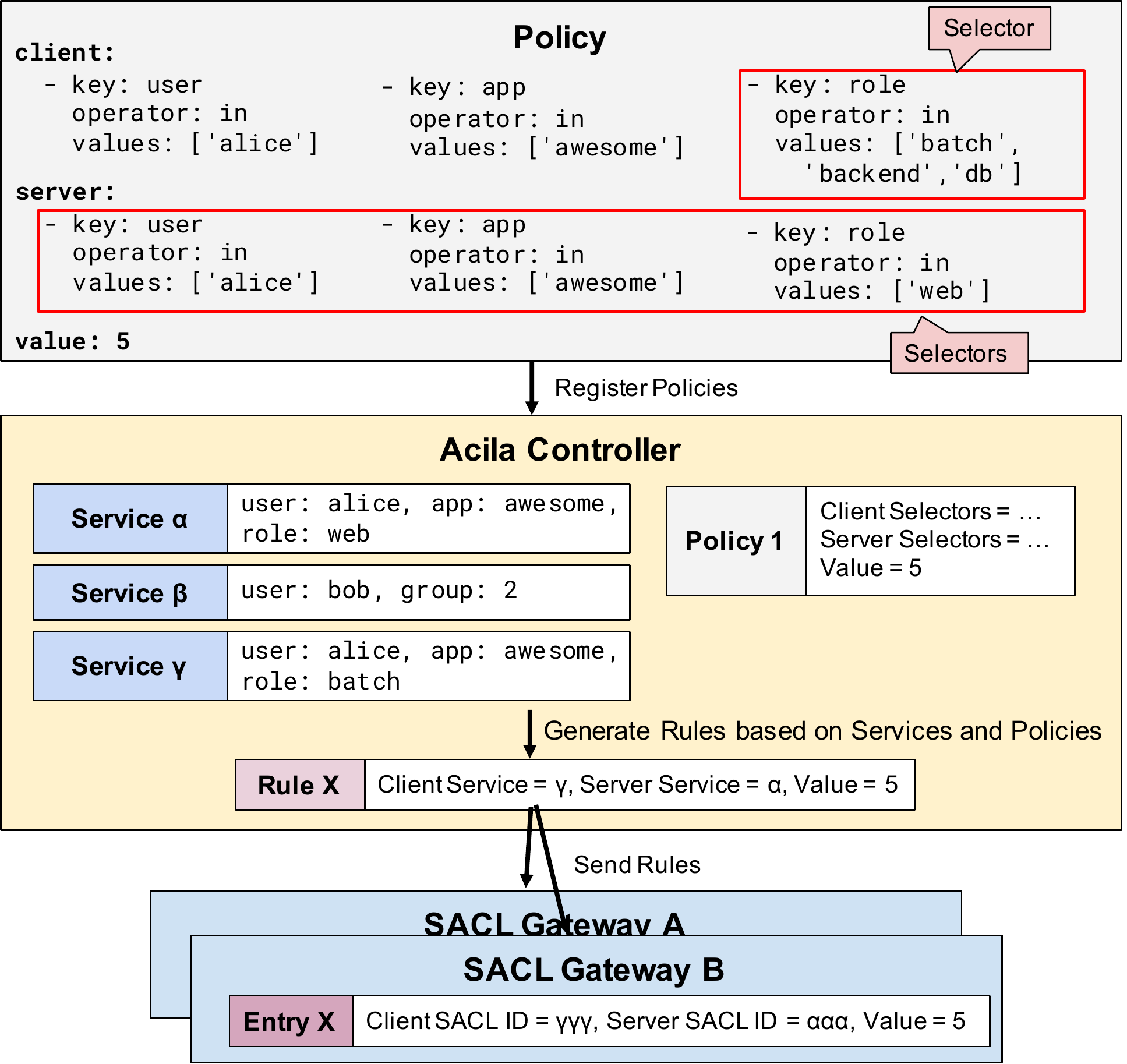}
		\caption{Policy management and distribution}
		\label{fig:policy}
	\end{center}
\end{figure}

In the following, we will describe the process inside Acila Controller. An overview and examples are shown in Figure~\ref{fig:policy}.

Selectors are used to select Services whose labels meet the conditions specified by the Selectors.
Each Selector consists of: Key, which is a key of label to be searched for, Operator which determines whether the value corresponding to the Key should exist (\texttt{in}) or not (\texttt{not\_in}), Values, which is a set of values that should or should not exist.
If there is more than one Selector, they are ANDed.

A Policy has two Selectors, one for the client-side and the other for the server-side.
When a Policy or a Service is registered or updated, Acila Controller finds a group of Services for both sides that match the Policy respectively.
Then it generates pairs of \texttt{(Client Service, Server Service, Value)} called \emph{Rule}, and distribute entries based on the Rules to the appropriate network devices.

Policies are also used to select the Server Workload required for the client-side SACL Gateway to put Server SACL IDs into packets.
Acila Controller distributes the information of Server Workloads so that each SACL Gateway identify Server Workloads where the Client Workloads under the SACL Gateway may send packets.

\subsection{Packet Filtering}

When a Client Workload initiates a session, we enforce policies at the SACL Gateway on the server-side based on the SACL ID in the packet.
Value is not used since Policy only needs to represent the client/server Services to be allowed to communicate.

SACL Gateways on the server-side (`SACL Gateway (server-side)' in Figure~\ref{fig:how-to-attach-sacl-id}) filter out the packets from the Client Workload to the Server Workload.
If this is done only on the client-side, all SACL Gateways on the client-side (`SACL Gateway (client-side)' in Figure~\ref{fig:how-to-attach-sacl-id}) need to filter out packets correctly in order to prevent undesired access to one Server Workload.
This may cause a delay until the policy is updated in the network.
In the proposed method, only SACL Gateways that take care of the Server Workload, which is expected to be smaller than the number of client-side SACL Gateways, have to update the rules so the delay is expected to be shorter.

Packets in the reverse direction, from the Server Workload to the Client Workload, are handled by Conntrack as in Section~\ref{sec:add_saclid}.
When the session has been already recorded in the Conntrack, the packet is allowed.

In addition, network devices on the network perform packet filtering.
In this case, the devices that performs packet filtering based on the SACL ID, such as a programmable switch, needs to be installed between the SACL Gateway on the client-side and the server-side.
This method is mainly for supporting connections to legacy systems that is not managed by the cloud orchestrators.

\subsection{Priority Control}

Priority control assigns a priority to packets based on the Services of both sides, and prioritizes packet processing on network devices.
In this case, Policy consists of Selectors for the client/server Services to set the priority and a Value which represents the priority to be set.
Acila Controller distributes entries consisting of \texttt{(Client SACL ID, Server SACL ID, Value (Priority))} to all devices that enforce priority control based on SACL ID\@.

\section{Implementation}

\label{sec:implementation}

\begin{figure}[ht]
	\begin{center}
		\includegraphics[width=\linewidth]{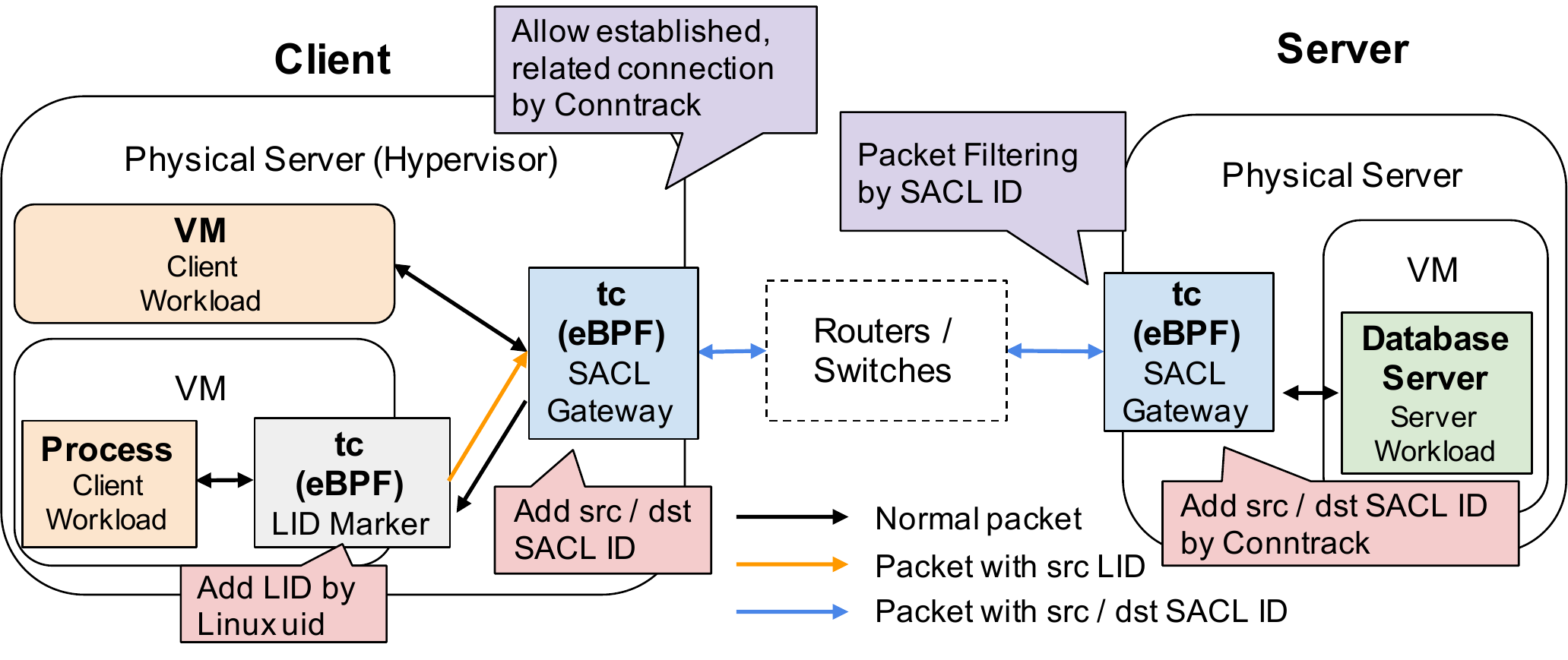}
		\caption{Overview of the data plane in the prototype system}
		\label{fig:acl_implementation}
	\end{center}
\end{figure}

We implemented a prototype system for packet filtering using the proposed method, including Service assignment to workloads, the addition of SACL IDs to packets. An overview is shown in Figure~\ref{fig:acl_implementation}.

We add SACL ID to packets between VMs and perform packet filtering.
Client Workload is a VM or a UNIX process in a VM and the Server Workload is a UNIX process in a VM\@.
When the Client Workload is a UNIX process in a VM, the LID Marker is installed in the VM\@.
SACL Gateway is installed in the hypervisor host of each VM\@.
SACL ID is 64 bits long.

Each VM has an IPv6 address and is connected to each other including the hypervisor hosts via Linux Bridge.

SACL Gateway is implemented on the hypervisor host using eBPF~\cite{eBPF}.
The eBPF program adds Client/Server SACL ID to packets from VMs and filters out the packets to VMs based on the SACL IDs.
It is attached to traffic control (tc) on the TAP interface connected to each VM\@.
Also, we implement a simple Conntrack mechanism in the server-side SACL Gateway to store correspondence between network based identifiers and a SACL ID\@.

LID is set in the Hop Limit field of the IPv6 header~\cite{IPV6} in the form of $LID \%128+ 100$.
Hop Limit is reset to the default in the SACL Gateway on the hypervisor host.
LID Marker uses the UID of UNIX process to identify the Client Workload.
LID Marker is implemented with eBPF, configured in tc on the interface which is the default gateway in the VM\@.

Acila Controller is implemented as a web server. For simplicity, we assume that the registering workload is divided into Client Workload and Server Workload in advance.

\section{Evaluation}

We evaluate the proposed method in terms of the number of entries by simulation and the performance of SACL Gateway.
We pick up priority control because, in order to perform priority control effectively, rules should be applied to all the network devices that packets pass through, and a larger number of entries are required compared to packet filtering and bandwidth control that can be applied only at SACL Gateway or network devices at the end.

\subsection{The Number and Update Amount of Entries}
\label{sec:eval_entries}
This section evaluates how much the number of entries and their update amount can be reduced by introducing the proposed method.

\textbf{Cloud Platform Assumption} We assume a 3-tier CLOS network~\cite{CLOS1}~\cite{CLOS2} (Leaf-Spine architecture~\cite{CLOS3}) as the network topology in the cloud platform.
This has been adopted in recent data center networks due to its ease of scale-out.
Each physical server in a server rack is connected to a ToR (Top of Rack) switch in each server rack.
Each ToR switch is connected to a leaf switch.
All leaf switches are connected to all spine switches, and vice versa.
We simulate the entries required for spine switches in addition to SACL Gateways since spine switches have entries covering all workloads or Services.

Here, each rack and physical server is denoted by $r_i$ and $m_i$, respectively.
A set of racks, a set of servers in the network, and a set of servers in rack $r_i$ are denoted as $R$, $M$, and $M_{r_i}$, respectively.

Multiple VMs running in a physical server, and multiple workloads are on a VM\@.
We denote each VM and each workload as $v_i$ and $w_i$ respectively.
A set of all workloads, a set of VMs in physical server $m_i$ and a set of workloads on VM $v_i$ are represented as $W$, $V_{m_i}$ and $W_{v_i}$, respectively.
The number of all workloads $|W|$ is calculated as
\begin{equation}
	|W| = \sum_{r_i \in R} \sum_{m_j \in M_{r_i}} \sum_{v_k \in V_{m_j}} |W_{v_k}| \nonumber %\label{eq:W}
\end{equation}

We assume that each workload is assigned an independent network based identifiers since we want to compare the number of entries in the case where both the conventional method using network based identifiers and the proposed method achieve the same priority control.

Each Service is denoted by $s_i$, and a set of all Services is denoted by $S$.
The set of Server Services whose connections from a Client Service $s_i$ have configured a priority is denoted by $\mathit{SS}_{s_i}$. 
The set of Client Services where a Server Service $s_i$ accept connections with priority-controlled packets is denoted as $\mathit{CS}_{s_i}$.

A workload belongs to one Service.
$W_{s_i}$ represents a set of workloads that belong to Service $s_i$.
$c_{w_i,w_j}$ shows the number of active connections between Client Workload $w_i$ and Server Workload $w_j$.

\subsubsection{Conventional Network-identifier Based Approach} 

With conventional approach, an additional Conntrack table will be required in order to differentiate whether a packet is from a client to a server or vice versa.
To avoid this differentiation, $\mathit{CS}_{s_i}$ and $\mathit{SS}_{s_i}$ are treated as the destination and source Services of packets instead of the Client and Server Services, respectively.
An entry for priority control should be configured for each workload, and consists of the network based identifiers of the source workload and the destination workload, and priority value.

Since each spine switch is connected to all leaf switches, the number of workloads under a spine switch is equal to $|W|$.
This value is the same among all spine switches.

We then discuss the number of entries required for a spine switch defined as $\mathit{el}$.
Since we cannot distinguish the direction of packets, from client to server or vice versa, we assume that a priority is set to packets from the destination Service to the source Service when a priority is set to packets from the source Service to the destination Service, in order to align the conditions with the proposed method.
In that case, a prioritized inter-workload communication is counted in both $\mathit{CS}_{s_i}$ and $\mathit{SS}_{s_i}$.
The number of entries required for a workload is the total number of workloads belonging to each destination Service that the workload makes connections to with priority configuration.
Therefore, $\mathit{el}$ is as follows.
\begin{equation}
	\mathit{el} =  \sum_{w_i \in W} \sum_{s_k \in \mathit{SS}_{s_j} \atop (W_{s_j} \ni w_i)} |W_{s_k}| \nonumber
\end{equation}

Next, we consider the number of entries to be updated in a spine switch when there is a change in workloads, Services, or the priority between Services. Note that `update' here includes the addition and removal of workloads and Services.

When a workload $w_i$ is created or deleted, the entries for the cases where the workload is at the source and at the destination are added or removed from a spine switch.
Therefore, the number of entries to be updated at a spine switch, $\mathit{elu}_{w_i}$, is
\begin{equation}
	\mathit{elu}_{w_i} = \sum_{s_k \in \mathit{SS}_{s_j}} |W_{s_k}| + \sum_{s_l \in \mathit{CS}_{s_j}} |W_{s_l}| \quad (W_{s_j} \ni w_i) \nonumber %\label{eq:el_per_p_sec}
\end{equation}

Creation or deletion of a Service is always accompanied by creation or deletion of workloads, and there is no addition or removal in entries caused by the Service.
Therefore, the number of updated entries per spine switch when Service $s_i$ is created or deleted, $\mathit{elu}_{s_i}$, is as follows.
\begin{eqnarray}
	\mathit{elu}_{s_i} & =& \sum_{w_i \in W_{s_i}} \mathit{elu}_{w_i} \nonumber
\end{eqnarray}

When one priority between Services is updated, priorities in both directions should be set.
For each direction, the number of updated entries per spine switch when the priority from Service $s_i$ to Service $s_j$ is updated,$\mathit{elu}_{{s_i}{s_j}}$, is as follows.
\begin{eqnarray}
	\mathit{elu}_{{s_i}{s_j}} &=& |W_{s_i}| * |W_{s_j}| \nonumber
\end{eqnarray}

\subsubsection{With Proposed Approach}

We assume that SACL Gateways are installed on physical servers and spine switches apply priority controls using SACL ID\@.
An entry for priority control on spine switches is in the form of (Client Service, Destination Service, Priority) and is required for each Service.

A Client Service $s_i$ requires entries for each Server Service ($\mathit{SS}_{s_i}$), so the number of entries required for a spine switch ($\mathit{es}$) is as follows.
\begin{eqnarray}
	\mathit{es} &=&  \sum_{s_i \in S} |\mathit{SS}_{s_i}| \nonumber
\end{eqnarray}

Suppose that a SACL Gateway $g_i$ is installed on a server $m_i$.
The number of elements in the set of workloads under SACL Gateway $g_i$, $W_{g_i}$, is as follows:
\begin{eqnarray}
	|W_{g_i}| = \sum_{v_j \in V_{m_i}} |W_{v_j}| \nonumber
\end{eqnarray}

SACL Gateway $g_i$ has the following three types of entries.

\textbf{Entries to add Client SACL ID} They are pairs of (IP address and LID (if any), SACL ID) for all Client Workloads under the SACL Gateway.
The number of entries, $\mathit{escc}_{g_i}$, is as follows.
\begin{eqnarray*}
      \mathit{escc}_{g_i} &=& |W_{g_i}| 
\end{eqnarray*}

\textbf{Entries to add Server SACL ID} They are pairs of the network based identifiers and the SACL ID of the Server Workload with which all Client Workloads under the SACL Gateway may communicate. Considering the worst-case in which all workloads in a physical server belong to different Services and all workloads of the Server Services do not overlap, the number of entries, $\mathit{escs}_{g_i}$, is as follows.
\begin{eqnarray*}
      \mathit{escs}_{g_i} &=& \sum_{w_j \in W_{g_i}} \sum_{s_l \in \mathit{SS}_{s_k} \atop (W_{s_k} \ni w_j)} |W_{s_l}| 
\end{eqnarray*}

\textbf{Entries to add SACL ID by Conntrack} An entry consists of two values: one is a pair of (Network based identifier of the Client Workload, Network based identifier of the Server Workload) and the other is a pair of both SACL IDs.
Let $c_{w_i}$ be the number of active connections of the Server Workload $w_i$ to the Client Workloads. $c_{w_i}$ is calculated as follows.
\begin{equation}
	c_{w_i} = \sum_{s_k \in \mathit{CS}_{s_j} \atop (w_i \in W_{s_j})} \sum_{w_l \in W_{s_k}} c_{{w_l},{w_i}} \nonumber
\end{equation}
Therefore, the number of entries of Conntrack, defined as $\mathit{ess}_{g_i}$, is as follows.
\begin{eqnarray*}
	\mathit{ess}_{g_i} &= &\sum_{w_j \in W_{g_i}} c_{w_j} \nonumber 
\end{eqnarray*}

The total number of entries in SACL Gateway $\mathit{es}_{g_i}$ is $\mathit{es}_{g_i} = \mathit{escc}_{g_i} + \mathit{escs}_{g_i} + ess_{g_i}$.

When one workload $w_i$ is created or deleted, there is no need to change the entries.
So, the number of updated entries in a spine switch in this case, $\mathit{esu}_{w_i}$, is as follows.
\begin{eqnarray}
	\mathit{esu}_{w_i} & =& 0 \nonumber
\end{eqnarray}

Creation or deletion of a Service $s_i$ requires adding or deleting both entries for the cases where the Service is Client Service and Server Service.
The number of updated entries in a spine switch in this case, $\mathit{esu}_{s_i}$, is as follows.
\begin{eqnarray}
	\mathit{esu}_{s_i} & =& |\mathit{SS}_{s_i}| + |\mathit{CS}_{s_i}| \nonumber
\end{eqnarray}

When one priority between Services is updated, only the entries related to the updated Services need to be changed.
So, let $s_i$ be the Client Service and $s_j$ be the Server Service whose priority is updated, and the number of updated entries per spine switch, $\mathit{esu}_{{s_i}{s_j}}$, is
\begin{eqnarray}
	\mathit{esu}_{{s_i}{s_j}} & =& 1 \nonumber
\end{eqnarray}

\subsubsection{Comparison}

We can confirm the number of entries in a spine switch is greatly reduced from the conventional approach $\mathit{el}$ to the proposed approach, $\mathit{es}$ as follows.
\begin{eqnarray*}
	\mathit{es} &=&  \sum_{s_i \in S} |\mathit{SS}_{s_i}| \\
	&=& \sum_{s_i \in S} \frac{|W_{s_i}|}{|W_{s_i}|} \sum_{s_k \in \mathit{SS}_{s_i}} \frac{|W_{s_k}|}{|W_{s_k}|} \\
	&=& \sum_{w_i \in W} \sum_{s_k \in \mathit{SS}_{s_j} \atop (W_{s_j} \ni w_i)}   \frac{|W_{s_k}|}{|W_{s_k}| |W_{s_j}|} < el
\end{eqnarray*}
This is because we can represent the combination of Client Workload $w_{l} \in W_{s_j}$ and Server Workload $w_{l} \in W_{s_j}$ by one entry.
Instead, servers need to maintain a certain number of entries.
This issue is discussed in Section~\ref{sec:saclgateway_performance}.

We can also see that the number of updated entries when a workload or a Service is added or deleted, or when a priority is changed because multiple workloads are usually in one Service.

\subsection{Performance of SACL Gateway}

\label{sec:saclgateway_performance}

SACL Gateway adds SACL IDs to packets in order to process packets based on SACL ID.
This means SACL Gateway has to perform additional processing that is not needed in conventional approach.

To evaluate the performance impact caused by SACL Gateway, we run a version of SACL Gateway that removed the packet filtering mechanism from the implementation explained in Section \ref{sec:implementation}.
We measure the change in RTT and throughput by the number of entries in SACL Gateway.

\begin{figure}[ht]
	\begin{center}
		\includegraphics[width=\linewidth]{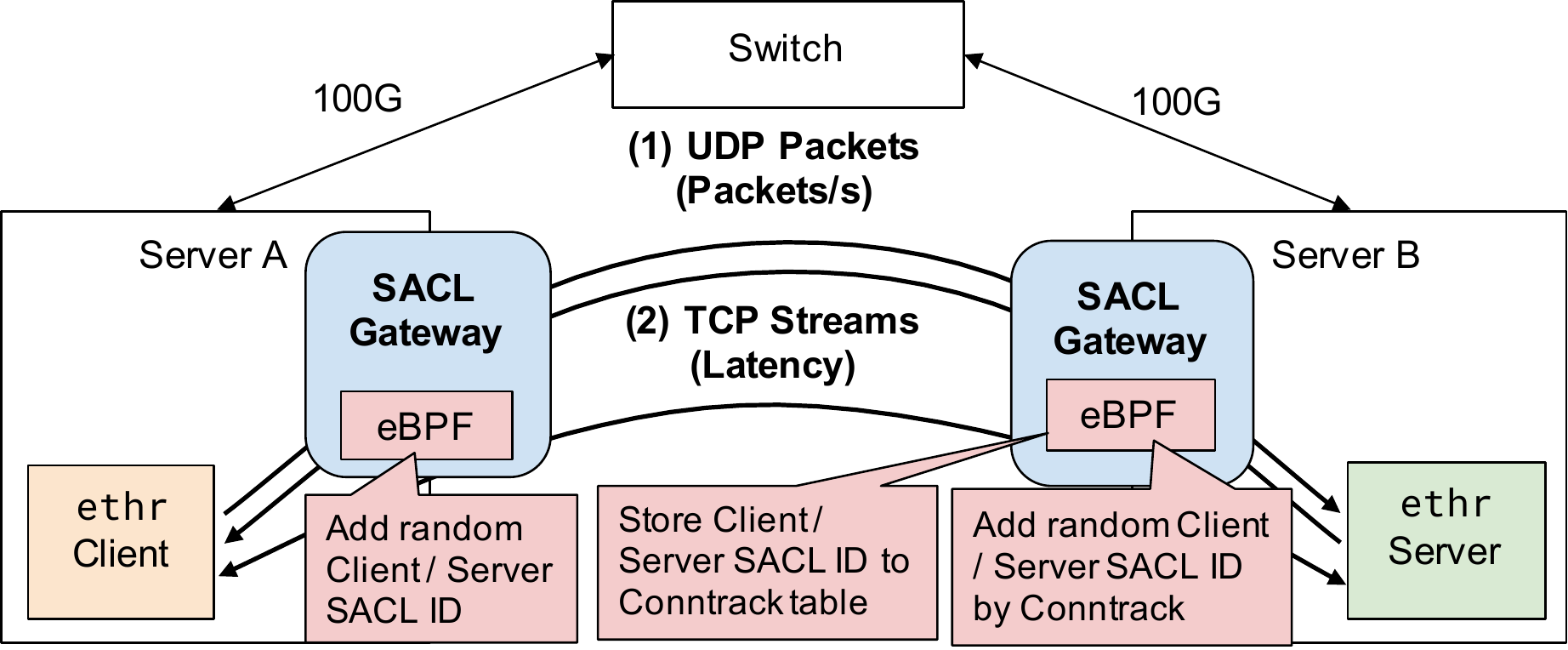}
		\caption{Overview of SACL Gateway performance evaluation}
		\label{fig:experiment}
	\end{center}
\end{figure}

Figure~\ref{fig:experiment} shows the overview of the environment conducting this evaluation.
Two servers, Server A and Server B, are connected via an Ethernet switch by 100GbE\@. SACL Gateway is configured on tc of the physical interface on both servers.
We set Server A as the client and Server B as the server.
We have measured the number of packets per second (pps, (1) in Figure~\ref{fig:experiment}) when massive UDP packets have sent from Server A or Server B, and the RTT when TCP connections have been established between Server A and Server B ((2) in Figure~\ref{fig:experiment}).

Each server has AMD EPYC 7282 (16 core, 32 thread), 128GB RAM, and Arch Linux (Linux 5.9.9-arch1-1) is installed.
Mellanox ConnectX-5 is installed as 100GbE NIC, and queue length for both tx and rx is set to 8192.

We have used Ethr~\cite{Ethr} 1.0.0 as a measurement program.
The number of simultaneous streams has been set to 16 and the measurement time to 60 seconds.
Server A has run Ethr as a client and vice versa.

For pps measurement, UDP packets with 1 byte of payloads were generated.
The measurement has been performed in two cases independently, when the SACL Gateway is on the client-side (Server A) and when it is on the server-side (Server B).
For RTT measurement, TCP connections has been used.

The SACL Gateway on Server A/B has been implemented with eBPF as described in Section \ref{sec:implementation}.
We have made several changes from the original implementation for the measurement.

\textbf{Removal of the packet filtering mechanism}
In this experiment, we have removed the packet filtering mechanism since we want to see the overhead of attaching and removing SACL IDs, not filtering packets.

\textbf{Randomized entry operations}
In Ethr, the number of clients is the same as the number of connections, the default value is 16, and the number of servers is one.
Since entries stored in the SACL Gateway are limited to those connections, it is insufficient for the measurement of a large number of entries that will be needed in cloud platforms.
Therefore, in this experiment, we reproduced the cost of searching/updating a large number of entries by randomizing the keys to search/update as follows.

Given that the number of entries in the evaluation scenario is $X$, we insert $X$ entries at the beginning so that the key of the $n$th entry is an unsigned integer $n$ and the value a randomized dummy SACL ID\@.
When the SACL Gateway is searching or updating entries (entry operations), the key is randomly selected: $(rand \% X)$ is used as the key, where $rand$ is a random 32-bit unsigned integer.
This lookup always hit.
Although randomized keys are gathered in bit space differently from the actual environment, there is no gap in the performance during entry operation because all the entry tables used in the eBPF program are a hash map.
The hashed keys should be distributed since the type of the hash maps is \texttt{BPF\_MAP\_TYPE\_HASH} which uses Jenkins hash~\cite{Jenkins}.

On the client-side SACL Gateway (Server A), different prefixes of the key in Conntrack are used among for initial data insertion, search and update so that any packet from the client to the server never hit Conntrack to simulate the cost of searching and updating entries.
Note that normally Conntrack is hit after the first packet in a connection, but not used in the experiment.
 This does not increase performance but may decrease compared to the actual environment.

Despite that it is common for the same key from the same network based identifiers to be used in one connection, the key is changed every packets in the experiment.
This drawback may decrease performance but not increase it because it leads to an increase in the cache miss rate and the number of newly inserted entries to Conntrack.

While a random value is used for the key, the same operations are performed for the IPv6 header such as parsing in order to measure the correct performance impacts.

The number of entries used in the experiment is based on a simple assumption about the values.
The assumption is that: all servers have 8 VMs, each VM has 16 workloads, every Service include 15 workloads, a Service communicates with 2 Services with priority, 1 connection is established at any time between workloads.
As the scale of the cloud grow, more and more entries will be installed, so we introduce a positive coefficient $\alpha$.
As a result, the number of entries for SACL Gateway is estimated as follows.
\begin{eqnarray*}
	\mathit{escc}_g &=& 1.3 * 10^2 \alpha  \\
	\mathit{escs}_g &=& 3.8 * 10^3 \alpha \\
	\mathit{ess}_g &=& 3.8 * 10^3  \alpha \\
\end{eqnarray*}
Note that the numbers introduced by the simple assumption can be ignored when $\alpha$ is large, and LID is not assumed here and the table for LID is not used.

The measurements have been made when $\alpha$ was 0.1, 0.5, 1, 2, 4, 8, 16 and 32 respectively and when no SACL Gateway has been installed as the baseline.

\subsubsection{Result}

\begin{figure}[ht]
	\begin{center}
		\includegraphics[width=\linewidth]{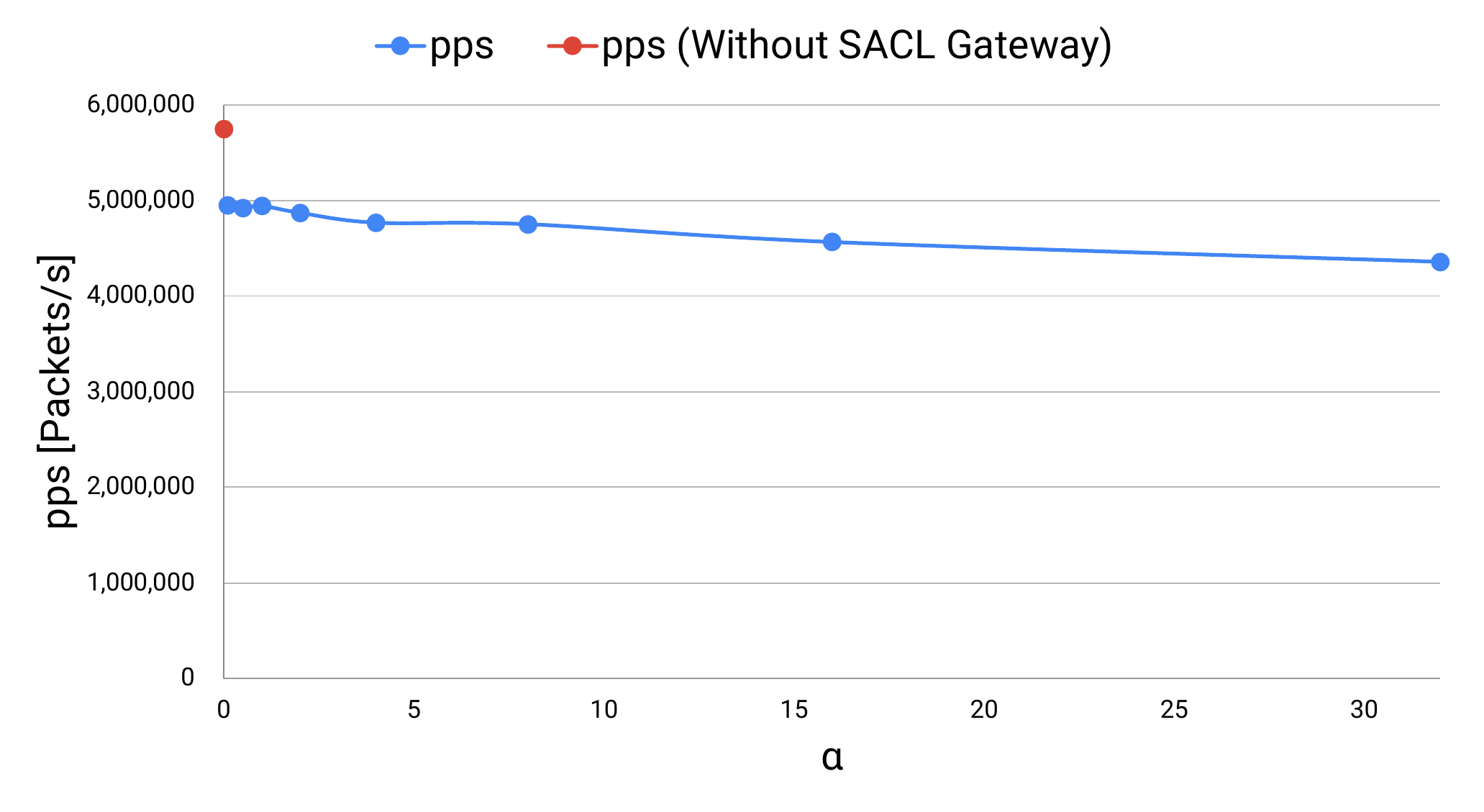}
		\caption{pps for different coefficient $\alpha$ for the number of entries, from a client to a server}
		\label{fig:pps}
	\end{center}
\end{figure}

Figure~\ref{fig:pps} shows the average pps for 60 seconds when packets are from the client to the server.

In the case where $\alpha = 1$,
the pps keeps $86 \%$ of the baseline.
The pps is decreasing when $\alpha$ becomes large.
Even when the number of entries is very large, $\alpha = 32$, the pps keeps $76 \%$ of the baseline.
On the other hand, even when the number of entries is small ($\alpha = 0.1$), the pps is reduced to $86 \%$ of the case without SACL Gateway.

\begin{figure}[ht]
	\begin{center}
		\includegraphics[width=\linewidth]{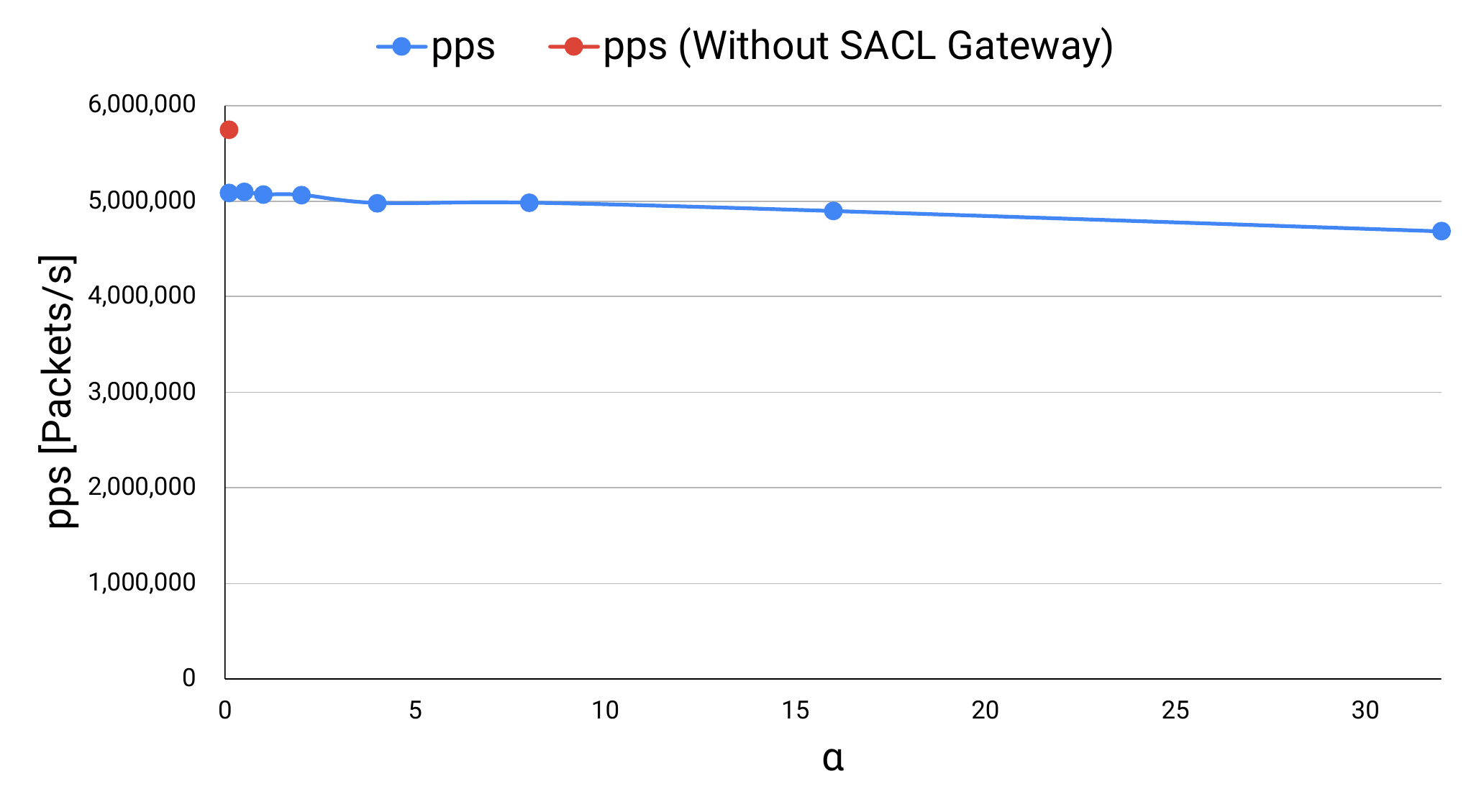}
		\caption{pps for different coefficient $\alpha$ for the number of entries, from a server to a client}
		\label{fig:reverse-pps}
	\end{center}
\end{figure}

The average pps for 60 seconds when packets are transmitted from the server to the client is shown in Figure~\ref{fig:reverse-pps}.

In the case where $\alpha = 1$, the pps keeps $88 \%$ of the baseline.
The pps is decreasing when $\alpha$ becomes large, but not as much as the case from the client to the server.
Even when the number of entries is very large, $\alpha = 32$, the pps also keeps $82 \%$ of the baseline.
On the other hand, even when the number of entries is small ($\alpha = 0.1$), the pps is also reduced to $88 \%$ of the case without SACL Gateway.

\begin{figure}[ht]
	\begin{center}
		\includegraphics[width=\linewidth]{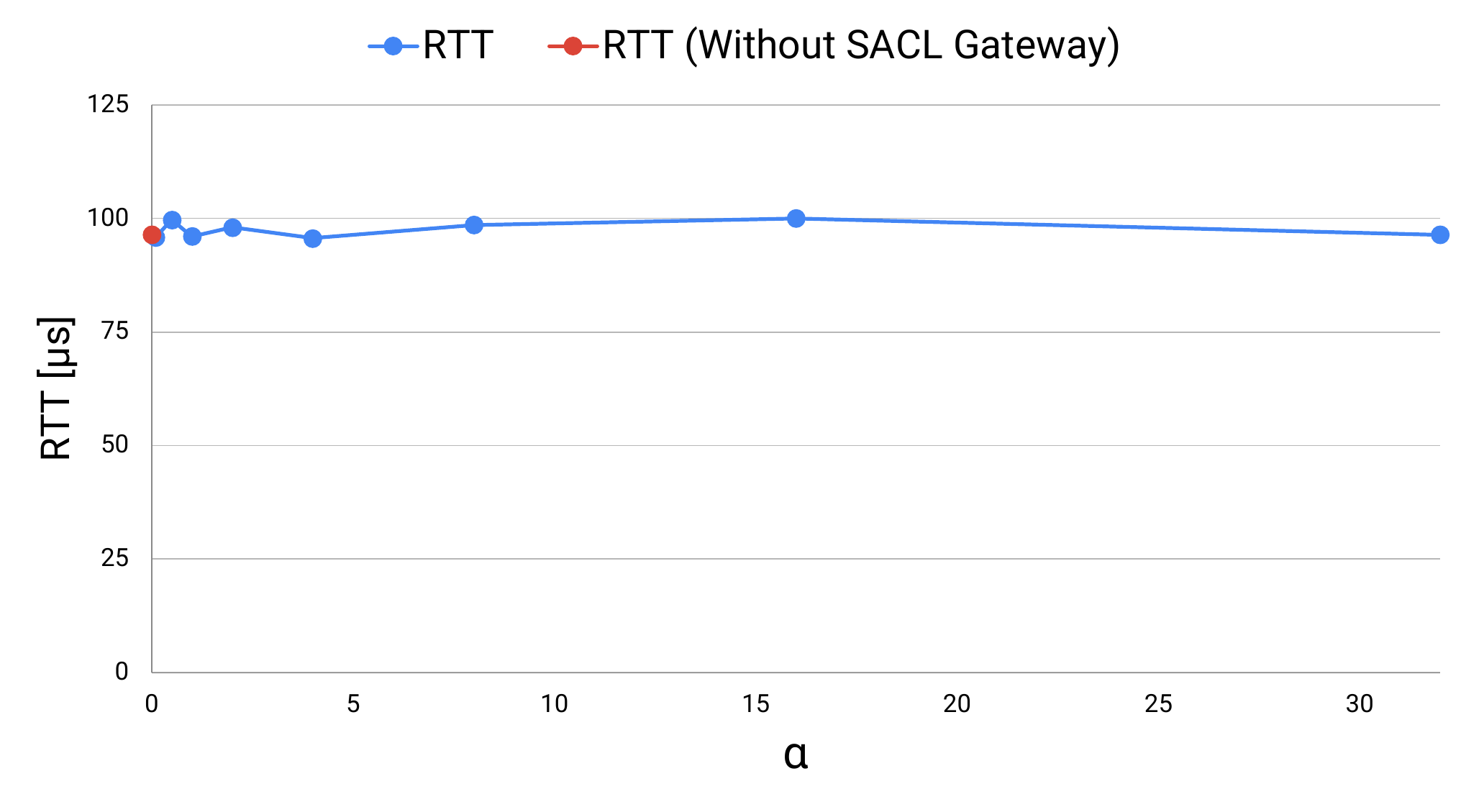}
		\caption{RTT between Server A and Server B for coefficient $\alpha$ for the number of entries}
		\label{fig:rtt}
	\end{center}
\end{figure}

The average RTT over 60 seconds for TCP connections from Server A to Server B is shown in Figure~\ref{fig:rtt}.
For $\alpha = 1$, the RTT is almost the same with the baseline.
We can also see that the RTT is almost stable when $\alpha$ becomes large.

\subsubsection{Discussion}

The performance degradation due to the introduction of the SACL Gateway with the expected number of entries was less than $15 \%$.
It was also found that the impact of the number of entries to pps is small.
Therefore, the performance does not decrease significantly even if the number of entries increases by an increase of workloads or introducing packet filtering.

The decrease in pps for the number of entries is smaller for packets from the client to the server than for packets from the server to the client is thought to be that the former requires three table searches (Conntrack, Client Workload, and Server Workload) while the latter requires only one (Conntrack).

When the number of entries is small, $\alpha = 0.1$, pps is $86 \%$ and $88 \%$ of each baseline, even though the search cost is considered to be low.
This should be due to the cost of parsing and expanding packets to insert SACL IDs at the SACL Gateway.
This may be improved by inserting SACL IDs during the packet assembly process using a kernel module, etc., instead of parsing and expanding the packet by eBPF after the packet has been assembled.

As for the RTT, there is no change in the presence of SACL Gateway or the number of entries, and the processing time of the SACL Gateway is so small that it can be ignored.
Moreover, in many cloud platforms, packets usually pass through more devices such as ToR switches and leaf/spine switches than in the experimental environment.
In such a platform, the baseline of RTT is expected to be higher.
Therefore, the impact of the processing time of the SACL Gateway to the RTT is almost negligible.

\section{Discussion}

\label{sec:kousatsu}

\subsection{Application to Larger Scale Cloud Platforms}

In very large cloud platforms having large scale networks, the following issues will appear.

\textbf{Increase in the Number of Entries} When a spine switch maintains the rules between all Services as it does now, the number of workloads and Services will increase, and the number of entries ($\mathit{es}$) will increase.
As a result, the number of entries may exceed the capacity of the switch.
In response to this, the following solutions can be considered.

\textit{Optimized Service generation}
In this paper, labels of a Service have a one-to-one correspondence with labels of a workload, but the labels may contain unnecessary information such as a serial number, which may increase the number of Services and entries.
It is possible to specify and limit the key of the label used to associating with the Service, or generate a Service with the appropriate keys to classify packets in all use cases introduced in the network.

\textit{Structured SACL ID}
Structuring SACL IDs is also a promising approach so that network devices can decide the actions to be performed by looking at a part of the SACL ID\@.
There are various ways to do this; assigning a common upper bit determined by a subset of labels for a Service, similar to entries of routing tables.
With structured SACL ID, the number of entries in a network device can be reduced by handling a group of Services together instead of entries for each Service.

\textbf{Increasing the Load of Acila}
Currently, all the information used is centrally managed by a single Acila Controller.
It may become difficult to manage them in a single instance, due to the increased load on the DB for example.
We can consider operating multiple Acila Controller instances such as on each data center.
There are some concerns about operating multiple Acila Controller instances across the data centers as follows.

\textit{Support for Policies} When Services are managed in a distributed manner, an Acila Controller instance cannot know which Service defined in another instance matches a policy.
There are some solutions, one is to centrally manage Services and their associated labels in one place and each Acila Controller instance converts them into rules by querying them to the centralized management entity.
Another one is to develop a mechanism that allows Acila Controllers to query each other for Services that match a policy.
The latter method has better scalability in terms of the load of Acila Controller, but the presence of many Acila Controllers may lead to degrade the performance by locking the database when updating a Policy locking problems, and so on.

\textit{Issuance of a common SACL ID} The same SACL ID must be used for the same Service in multiple Acila Controllers.
This can be done by managing the issuance of SACL IDs in one place as described above.
Another solution is to generate a common SACL ID in a distributed manner.
We can use the hash value from labels as SACL ID with a hash function having sufficient collision resistance.
However, there is a small possibility of collision, and we cannot simply use the hash function when SACL ID is structured.

\textbf{Exhaustion of SACL ID in Bit Space} While this research does not specify the length of SACL ID in the packet format, when the length is small, there is a possibility that the SACL ID bit space will run out as the scale of the cloud platform grows and the number of Services increases.
On the other hand, if the length is unnecessarily large, the packet length will become longer and the goodput will be decreased.

\subsection{Precise Workload-Service Association}
Acila relies on network based identifiers to associate a workload with a Service, and SACL Gateways must have a large number of entries to do this.
SACL Gateways should be richer information for this association, such as an interface where a workload is connected, and use of such information may reduce the number of entries.
Applications running on workloads have more data, and such data is also useful for the association.
For example, designing a new socket option so that applications pass some clues to identify a Service, and the kernel or other modules use the clues to associate with a Service.

\subsection{Protocol}

The protocol for adding SACL ID uses IPv6 Hop-by-Hop Options, this means that IPv4 packets are not currently supported and needs to be encapsulated in IPv6 packets.
Header fields in the packet format itself is independent of IPv6 addresses, so we can easily design the same options in IPv4 packets, but the compatibility with network devices that do not support SACL ID should be carefully assessed.

\section{Conclusion}

Use of network based identifiers, IP addresses and port numbers, are inappropriate for classify source and destination hosts of packets in the cloud platforms when identity of such hosts are required to configure the classification rules.
This is due to the fact that the cloud elastically creates and deletes workloads in servers whose resources are available according to the load on each system, and that network based identifiers are assigned sorely by the location of the servers.
Therefore, the number of entries for such classification in network devices is large, and the entries should be frequently updated.
To solve these problems, we proposed Acila, a system that extracts the information related to workloads, and reduces the number and update frequency of the entries by handling multiple workloads together as Services.
The system assigns identifier of Services (SACL ID) to packets between workloads so that network devices can identify Services that the source and destination of packets belong to.
We also proposed applications to packet filtering and priority control using Acila.
Through evaluation, we confirmed the effectiveness of Acila by evaluating the number of entries required for simulated priority control.
We also benchmarked the performance of our SACL Gateway implementation and confirmed that there was no significant overhead.

Future works include extending Acila to support more large scale cloud platforms, which hosts tens of thousands of server racks by structuring SACL IDs and optimizing Service generation methods.

\nocite{*}

\bibliographystyle{ACM-Reference-Format}
\bibliography{main}

\end{document}